\documentclass[aps,pra, twocolumn,nofootinbib, showpacs]{revtex4-1} 
\usepackage[unicode=true,pdfusetitle, bookmarks=true,bookmarksnumbered=false,bookmarksopen=false, breaklinks=false,pdfborder={0 0 0},backref=false,colorlinks=false] {hyperref}
\hypersetup{ colorlinks,linkcolor=myurlcolor,citecolor=myurlcolor,urlcolor=myurlcolor}
\usepackage{braket,colortbl,amsthm,amsmath,cleveref,amssymb,txfonts}\definecolor{myurlcolor}{rgb}{0,0,0.7}
\usepackage{graphics,graphicx}
\usepackage{color}
\usepackage{amssymb}
\usepackage{amsthm}
\usepackage{amsfonts}
\usepackage{float}
\usepackage{graphicx}
\usepackage[format = plain,labelfont = bf,up, textfont = normal , up, justification =raggedright, singlelinecheck =false]{caption}
\usepackage{subcaption}
\usepackage{tabularx}
\usepackage{amsmath}
\usepackage{braket}

\usepackage{graphicx}
\usepackage[utf8x]{inputenc}
\usepackage{color,soul}
\usepackage{amsmath}
\usepackage{braket}
\usepackage{latexsym}
\usepackage{amssymb}
\usepackage{amsthm}
\usepackage{xcolor}
\usepackage{bm}
\usepackage{graphics,epstopdf}
\usepackage{color}\usepackage{amsmath}
\usepackage{enumitem}
\usepackage{fmtcount}
\usepackage{booktabs}
\usepackage{csquotes}
\usepackage{epsfig}
\usepackage[normalem]{ulem}
\theoremstyle{plain}

\def\bea{\begin{eqnarray}}
	\def\eea{\end{eqnarray}}
\def\ba{\begin{array}}
	\def\ea{\end{array}}

\def\beq{\begin{equation}}
	\def\eeq{\end{equation}}

\begin{document}
	\title{Measurement-device-independent nonlinear entanglement witnesses}

	\author{Kornikar Sen, Chirag Srivastava, Ujjwal Sen}
	
	\affiliation{Harish-Chandra Research Institute, HBNI, Chhatnag Road, Jhunsi, Allahabad 211 019, India}
	\begin{abstract}
		Entanglement witnesses are one of the most effective methods to detect entanglement. It is known that nonlinear entanglement witnesses  provide better entanglement detection than their linear counterparts, in that the former detect a strictly larger subset of entangled states than the latter. Whether linear or nonlinear, the method is measurement-device dependent, so that imperfect measurements may cause false certification of entanglement in a shared state. Measurement-device-independent entanglement witnesses  provide an escape from such measurement dependence of the entanglement detection for linear entanglement witnesses.  Here we present measurement-device-independent nonlinear entanglement witnesses  for non-positive partial transpose entangled states as well as for bound entangled states with positive partial transpose. The constructed 
		measurement-device-independent nonlinear entanglement witnesses
certify the entanglement of the same sets of entangled states as their device-dependent parents do, and therefore are better than the linear entanglement witnesses, device-independent or otherwise. 
	\end{abstract}
	\maketitle
	
	\section{Introduction}
	\label{sec1}
	Entanglement \cite{Hor'09,Guhne'09,Das'19} is one of the key nonclassical concepts of  quantum theory. It is a fuel to various quantum information processing and computational tasks.
	Therefore, determination of the presence of entanglement in a given state is a crucial task.
	A necessary and sufficient criterion to guarantee entanglement is still to be found for arbitrary-dimensional systems.  But methods like the positive partial transpose (PPT) criterion \cite{Peres'96,Hor'96}, the range criterion \cite{Hor'97}, and the entanglement witnesses (EWs) \cite{Hor'96,Terhal'00,Bruss'02} can be applied in some cases to ensure entanglement in a quantum state. Several of the methods have also been implemented in  laboratories (see e.g. \cite{Di}).
	Specifically, the method of entanglement witnesses has been widely studied over the past years and is also considered to be one of the most viable ones to detect entanglement experimentally. The formulation of EWs is based on the ideas of the Hahn-Banach  separation theorem on normed linear spaces \cite{Simmons}. Since the separable states form a convex and closed set,  the Hahn-Banach theorem  guarantees the existence of inequalities on the set of quantum states which can separate some  entangled states from the set of all separable states. Therefore, inequalities can be constructed, whose violations guarantee entanglement. The Bell inequalities \cite{Bell} are particular examples of EWs. 
	
	An inequality to detect entanglement consists of expectations of certain operators which can be experimentally obtained using local measurements. This implies that the correctness of entanglement detection via an EW relies on that of the measurements being performed. And certain wrong measurements can even result in a wrong certification of entanglement \cite{See'01,Skwara,See'07,Morod,Bancal,Rosset}.
	Violation of a Bell inequality does guarantee entanglement independent of the measurements. But a Bell inequality typically certifies a lower volume of entangled states compared to the optimal EWs \cite{Werner'89,Scar,Bar,Las,De,Hyl,Methot}. To avoid the dependence on correctness of measurements and yet to maintain the optimality of EWs, the method of measurement-device-independent entanglement witnesses (MDI-EWs) was formulated \cite{Cyril}. It is based on a semi-quantum nonlocal game where all entangled states give a higher payoff compared to separable states \cite{Buscemi}. In contrast to the usual Bell inequality violation scenario or the standard EWs one, where the measurements are used as inputs to construct the inequality, a set of quantum states are used as inputs to evaluate a payoff function for any state in case of the semi-quantum nonlocal games. For other works on MDI-EWs, see \cite{Mallick,Abiuso,Sriv,Sen}. The measurement-device-independent scenario is also implemented in various other contexts like quantum  key  distribution \cite{Lo},  verification  of  quantum  steering \cite{Caval}, etc.
	Typically, inequalities considered for constructing EWs are linear, but there also exist several works on nonlinear inequalities to detect entanglement \cite{Guhne'06,Sen'19,Uffink}. In particular, nonlinear improvements over linear EWs, termed as nonlinear entanglement witnesses (NEWs), are studied in \cite{Guhne'06}. The idea is to use the nonlinearity of these inequalities to catch more entangled states. There also exist experimental implementations of NEWs \cite{Ant}.
	In this article, we present  measurement-device-independent nonlinear entanglement witnesses (MDI-NEWs), made out of nonlinear entanglement witnesses. They can detect the same entangled states as the corresponding NEWs can detect, but measurement-device-independently. Also, they provide improvement over MDI-EWs, i.e., they detect more entangled states compared to MDI-EWs.
	We construct MDI-NEWs for non-positive partial transpose (NPPT) entangled states and PPT bound entangled states. For bipartite systems, bound entangled states are those which cannot be distilled to singlets using local quantum operations and classical communication (LOCC),  even in the asymptotic limit \cite{Hor'97}.
	
	The rest of the paper is organized as follows. In Sec. \ref{sec2}, the concepts of nonlinear entanglement witnesses and  measurement-device-independent entanglement witnesses are briefly discussed. In Sec. \ref{sec3}, we show that MDI-NEWs can be created out of NEWs for NPPT entangled states, and in Sec. \ref{sec4}, MDI-NEWs for (PPT) bound entangled states are constructed. We summarize our results in Sec. \ref{sec5}.
	
	
	\section{Gathering the tools}
	\label{sec2}
	In this section, we will give  short overviews of linear and nonlinear EWs, and MDI-EWs. EWs were introduced based on the Hahn-Banach theorem \cite{Simmons}. The theorem states that 
	if we consider a closed and convex set, \(C\), inside a normed linear space $L$, and \(x\) is an element of \(L\) that is not in \(C\), 
	then there exists a continuous functional $f:L\rightarrow \mathbb{R}$ such that $f(c)\leq r < f(x)$ for all $c\in C$, where $r\in \mathbb{R}$. 
		The set of density matrices on a Hilbert space is part of a normed linear space, and the set of all separable states, say $S$, on a tensor-product Hilbert space, say $\mathcal{H}_A\otimes\mathcal{H}_B$, is a closed and convex set.
		According to the Hahn-Banach theorem, we can therefore always define a hyperplane which can distinguish between the set $S$ and an entangled state, say $\rho$, on the same Hilbert space, $\mathcal{H}_A\otimes\mathcal{H}_B$. Moreover, the set of entangled states is open, therefore that hyperplane will also be able to detect at least a small neighbourhood of the state $\rho$ as entangled. Using these concepts, the witness operator was defined as follows. An entanglement witness, $W$, is a hermitian operator which satisfies the following properties:
	\begin{enumerate}
		\item 
		$\text{Tr}(W\sigma)\geq0$ for all $\sigma\in S$.
		\item 
		There exists at least one $\rho\notin S$ such that $\text{Tr}(W\rho)<0$.
	\end{enumerate}
	Corresponding to every entangled state, $\rho$, there exists at least one witness operator, $W$, such that $\text{Tr}(W\rho)<0$ \cite{Hor'96, Terhal'00, Bruss'02}.
	
	Let $\rho_\phi$ be a non-positive partial transpose (NPPT) entangled state shared between two parties, \(A\) and \(B\), and let the corresponding Hilbert space be \(\mathcal{H}_A \otimes \mathcal{H}_B\). I.e., $\rho_\phi^{T_B}$ has at least one negative eigenvalue, where $T_B$ denotes transposition over the second party, $B$. Let the eigenvector corresponding to one of $\rho_\phi$'s negative eigenvalues be denoted by $|\phi\rangle$. Then the expectation value of the hermitian operator
	\begin{equation}
		W_\phi=|\phi\rangle\langle\phi|^{T_B} \label{eq5}
	\end{equation} is positive for all separable states, and is negative for the entangled state $\rho_\phi$. $W_\phi$ is an example of an EW \cite{Guhne'09,Terhal'00,Bruss'02}.

	In the next two subsections, we will briefly review two different type of functionals which can also be used for detection of entanglement and have some advantages over the one 
	discussed just now. 
	\subsection{Nonlinear entanglement witnesses}
	A single linear entanglement witness operator  is incapable of detecting all entangled states. If we ``bend'' the linear operator ``towards negativity'' in such a way that it still remains positive for all separable states, the resulting nonlinear functional will detect more entangled states than its linear parent. Such functionals are called NEWs. We can form a NEW, $F_\phi$, using the witness operator $W_\phi$ in the following way \cite{Guhne'06}:
	\begin{eqnarray}
		F_\phi(\rho)=\left<W_\phi \right>-\frac{1}{s(X)}&&\left< X^{T_B}\right>\left<\left(X^{T_B}\right)^\dagger\right> \nonumber\\
		=\left<W_\phi \right>-\frac{1}{s(X)}&&\left(\left<H_\phi\right>^2+\langle A_\phi\rangle ^2\right)\nonumber,\\&&\text{where }X^{T_B}=H_\phi+iA_\phi. \label{eq1}
	\end{eqnarray}
	Here $X=|\phi\rangle\langle \psi|$, where $|\psi\rangle$ is any arbitrary state and $s(X)$ is the square of the largest Schmidt coefficient of $|\psi\rangle$. $H_\phi$ and $A_\phi$ are taken to be hermitian operators. The last equality in the above equation is written using the fact that any operator can be written as a sum of hermitian and antihermitian operators. The expectation values in Eq. \eqref{eq1} are taken over the state $\rho$. It can be easily verified that $F_\phi(\sigma)$ is non-negative for all $\sigma\in S$. Also, since $\left(\left<H_\phi\right>^2+\langle A_\phi\rangle ^2\right)/s(X)$, which is a positive term, is subtracted from $\left<W_\phi \right>$, 
	$F_\phi$ will successfully detect more entangled states than $W_\phi$.
	\subsection{Measurement-device-independent entanglement witnesses}
	EWs provide one of the few methods which can detect entangled states without the need for tomography of the whole state, leading its importance in theoretical as well as experimental studies in quantum devices. 
	But to know if a state is entangled or not, the experimentalist
 will	need to measure the expectation value of the witness operator. Herein lies a crucial drawback of this method: 
 If the measurement operators get altered without the knowledge of the experimentalist, then 
 errors may occur in the corresponding expectation values, and thus a separable state may appear as entangled. MDI-EWs provide a way to avoid such a scenario \cite{Cyril,Buscemi}. 
	
	Let Alice and Bob share a state $\rho_{AB}$, which acts on a joint Hilbert space $\mathcal{H}_A\otimes\mathcal{H}_B$, of local dimensions \(d_A\) and \(d_B\). They want to find out if $\rho_{AB}$ is entangled. Since density operators span the space of hermitian operators, any EW operator, $W$, acting on $\mathcal{H}_A\otimes\mathcal{H}_B$, can be decomposed in terms of density operators, i.e, say $W=\sum_{s,t}\alpha_{st}\tau_s^T\otimes\omega_t^T$. Here $\{\tau_s\}_s$ and $\{\omega_t\}_t$ are sets of density matrices acting on the Hilbert spaces $\mathcal{H}_A$ and $\mathcal{H}_B$ respectively, and $\{\alpha_{st}\}_{s,t}$ is a set of real numbers. Further, let the set of states $\{\tau_s\}_s$ ($\{\omega_t\}_t$) be provided to Alice (Bob). We assume that Alice (Bob) can perform a joint dichotomic positive operator-valued measurement (POVM), $\{A_0,~A_1\}$ ($\{B_0,~B_1\}$), on a randomly chosen state $\tau_s$ ($\omega_t$) from the given set of states $\{\tau_s\}$ ($\{\omega_t\}$) and her (his) part of the state $\rho_{AB}$.
	Let $A_1$ and $B_1$ be denoted by outcome `1'.
	Let the probability that both Alice and Bob  will get 1 as output be denoted as $P(1,1|\tau_s,\omega_t)$. Then the MDI-EW is defined by using the quantity \cite{Cyril}, 
	\begin{equation*}
		I(P)=\sum_{s,t}\alpha_{st}P(1,1|\tau_s,\omega_t).
	\end{equation*}
	It can be shown that $I(P)$ is positive for all separable states, whatever be the joint POVM used by Alice or Bob. Moreover, if Alice and Bob choose $A_1$ and $B_1$ as the projectors of the maximally entangled state $|\Phi^+\rangle=\frac{1}{\sqrt{d_{A(B)}}}\sum_{i=1}^{d_{A(B)}}|ii\rangle$, then $I(P)$ reduces to the expectation value of $W$ divided by $d_A d_B$. Thus $I(P)$ can detect all the entangled states which can be detected using the expectation value of $W$. Even if the dichotomic measurement is kept arbitrary, $I(P)$ will never certify a separable state as entangled.    
	\section{MDI Nonlinear Entanglement witnesses}
	\label{sec3}
	It can be seen from Eq. \eqref{eq1} that to find out if a state is entangled using the NEW, one needs to measure expectation values of the three operators, $W_\phi$, $H_\phi$, and $A_\phi$. If the measurement process is not 
	reliable and the measurement operators have the risk to get replaced by some different and unknown operators, then
	the value obtained for \(F_\phi (\rho)\) will be erroneous. And as a result, the experimentalist 
	may misidentify a separable state as entangled. 
	Such misidentification can be avoided by using the MDI-NEW that we introduce now.
	
	
	We can  decompose $W_\phi$, $H_\phi$, and $A_\phi$, of Eq. \eqref{eq1}, in terms of two sets of density operators, say $\{\tau_s\}$ and $\{\omega_t\}$, in the following way:
	\begin{eqnarray*}
		W_\phi&=&\sum_{s,t}\alpha_{st}\tau_s^T\otimes\omega_t^T,\\
		H_\phi&=&\sum_{s,t}\beta_{st}\tau_s^T\otimes\omega_t^T,\\
		\text{and }A_\phi&=&\sum_{s,t}\gamma_{st}\tau_s^T\otimes\omega_t^T.	
	\end{eqnarray*} 
	Just like in the case of MDI-EWs, let us assume that  Alice (Bob) has the set of states $\{\tau_s\}$ ($\{\omega_t\}$) and can apply a joint dichotomic POVM, $\{A_0,~A_1\}$ ($\{B_0,~B_1\}$) on any state $\tau_s$ ($\omega_t$) and her (his) part of the shared state $\rho_{AB}$. 
	However, different from the case of MDI-EWs,  Alice and Bob have, in addition, been provided the maximally mixed states $m_A$ and $m_B$ respectively, that are maximally mixed in their respective Hilbert spaces, which they can also use instead of $\tau_s$ and $\omega_t$ in the joint measurements. Then the MDI-NEW can be defined by using the quantity,
	\begin{widetext}
	\begin{equation}
	N_\phi(P)=\sum_{s,t}\alpha_{s,t}P(1,1|\tau_s,\omega_t)-\frac{1}{s(X) d_A d_B P(1,1|m_A,m_B)}
	\left[\left(\sum_{s,t}\beta_{s,t}P(1,1|\tau_s,\omega_t)\right)^2+\left(\sum_{s,t}\gamma_{s,t}P(1,1|\tau_s,\omega_t)\right)^2\right]. \label{eq2}
	\end{equation} 
	\end{widetext}
	We claim the following: 
	\begin{enumerate}
		\item 
		$N_\phi$ is positive for all separable states. This positivity is independent of the applied POVMs.
		\item 
		$N_\phi$ can detect all  entangled states that can be detected using $F_\phi$ of Eq. \eqref{eq1}, when Alice and Bob use a particular joint POVM.
	\end{enumerate}
	To prove the first part, let us suppose that Alice and Bob share a separable state, $\sigma_{AB}=\sum_i p_i \sigma^A_i\otimes\sigma^B_i$, and
	the composite measurement operator corresponding to output (1,1) is $A_1\otimes B_1$. The maximally mixed states can be written as $m_A=I_A/d_A$ and $m_A=I_B/d_B$, where $I_A$ and $I_B$ are identity operators of the corresponding Hilbert spaces. Then we have
	\begin{eqnarray}
		P_\sigma(1,1|\tau_s,\omega_t)&=& \text{Tr}[(A_1\otimes B_1)(\tau_s\otimes\sigma_{AB}\otimes\omega_t)]
		\nonumber\\&=&\sum_i p_i \text{Tr}[(A_1^i\otimes B_1^i)(\tau_s^T\otimes\omega_t^T)] \nonumber,\\
		\text{and }P_\sigma(1,1|m_A,m_B)&=&\text{Tr}[(A_1\otimes B_1)(m_A\otimes\sigma_{AB}\otimes m_B)]\nonumber\\&=&\sum_i p_i \text{Tr}[A_1^i\otimes B_1^i]/(d_Ad_B).\label{eq3}
	\end{eqnarray} 
	Here, $A_1^i=\left(\text{Tr}_A[A_1(I_A\otimes\sigma^A_i)]\right)^T$ and $B_1^i=\left(\text{Tr}_B[B_1(\sigma^B_i\otimes I_B)]\right)^T$ are effective POVMs, where the partial traces in $A_1^i$ and $B_1^i$ are taken over the state spaces of $\sigma^{A}_i$ and $\sigma^{B}_i$ respectively. Let us define
	\begin{equation}
		Q=\left({\sum_i p_i A_1^i\otimes B_1^i}\right)/K\text{, where }K=\text{Tr}\left[\sum_i p_i  A_1^i\otimes B_1^i\right]. \label{eq4}
	\end{equation}  The operator $Q$ is hermitian, positive, as well as of unit trace.     
	Substituting $P_\sigma(1,1|\tau_s,\omega_t)$ and $P_\sigma(1,1|m_A,m_B)$ from Eq. \eqref{eq3} into Eq. \eqref{eq2} and using Eq. \eqref{eq4}, we get
	\begin{eqnarray*}
		N_\phi(P_\sigma)&=&K\text{Tr}[Q W_\phi]-\frac{1}{s(X)}\left(K\text{Tr}[QH_\phi]^2+K\text{Tr}[QA_\phi]^2\right)\\
		&=&KF_\phi(Q).
	\end{eqnarray*}
	In the last equality above, we have used Eq. \eqref{eq1} and the fact that $Q$ satisfies all the properties of a valid quantum state. Also from Eq. \eqref{eq4} it can be seen that $Q$ is a separable state. Thus $F_\phi(Q)\geq 0$. Since, $A_1^i$ and $B_1^i$ are positive operators, $K$ is also positive. Hence we have $N_\phi(P_\sigma)\geq0$ for all $\sigma_{AB}\in S$, and for arbitrary $A_1$ and $B_1$.
	
	Suppose now that the joint POVMs outcomes $A_1$ and $B_1$ are projections on the maximally entangled states, $|\Phi^+_A\rangle=\frac{1}{\sqrt{d_A}}\sum_{i=1}^{d_A}|i\rangle\otimes|i\rangle$ and $|\Phi^+_B\rangle=\frac{1}{\sqrt{d_B}}\sum_{j=1}^{d_B}|j\rangle\otimes|j\rangle$. Now the shared state between Alice and Bob is $\rho_{AB}$, which may not be separable. Then the probabilities with both the outputs as 1 are
	\begin{eqnarray}
		P_\rho(1,1|\tau_s,\omega_t)&=&\text{Tr}[(|\Phi^+_A\rangle\langle\Phi^+_A|\otimes|\Phi^+_B\rangle\langle\Phi^+_B|)(\tau_s\otimes\rho_{AB}\otimes\omega_t)]\nonumber\\
		&=&\text{Tr}[(\tau_s^T\otimes\omega_t^T)\rho_{AB}]/(d_Ad_B)\text{, and }\nonumber\\
		P_\rho(1,1|m_A,m_B)&=&\text{Tr}[(|\Phi^+_A\rangle\langle\Phi^+_A|\otimes|\Phi^+_B\rangle\langle\Phi^+_B|)(m_A\otimes\rho_{AB}\otimes m_B)]\nonumber\\&=&1/(d_Ad_B)^2.\nonumber
	\end{eqnarray}
	Hence from Eq. \eqref{eq2} we have
	\begin{equation}
		N_\phi(P_\rho)=\frac{\langle W_\phi\rangle}{d_Ad_B}-\frac{1}{s(X)}\left(\frac{\langle H_\phi\rangle^2}{d_Ad_B}+\frac{\langle A_\phi\rangle^2}{d_Ad_B}\right)=\frac{F_\phi(\rho_{AB})}{d_Ad_B}.
	\end{equation}  
	Thus $N_\phi(P_\rho)$ is negative for the states for which $F_\phi(\rho_{AB})$ is negative. This proves the second point.
	
	We see that $N(P)$ is  negative only for entangled states, and therefore it will not mistake a separable state as entangled because of wrong measurements. Thus, we can say that  $N(P)$ can be used as an MDI-NEW operator.
	
	$W_\phi$ is defined to detect the NPPT state $\rho_\phi$, and so it is plausible that $F_\phi$ and \(N_\phi\) will also be able to detect mostly NPPT states.
		In the succeeding section, we present a MDI-NEW which can be used to detect PPT entangled states. However, before that, we present an example below.

	\vspace{2 mm}

\noindent	\textbf{Example:} Werner states~\cite{Werner'89} form a well-known family of two-qubit states, and are defined as  $\rho_\nu=\nu \ket{\psi^-}\bra{\psi^-}+\frac{1-\nu}{4}I_4$, where $I_4$ is the identity operator on the two-qubit Hilbert-space and $\ket{\psi^-}=\frac{\ket{01}-\ket{10}}{\sqrt{2}}$. $\rho_\nu$ is separable for $\frac{-1}{3}\leq \nu \leq \frac{1}{3}$, entangled for $1\geq\nu>\frac{1}{3}$. The set of entangled Werner states can be easily detected by using the PPT criterion. The operator $\rho_\nu^{T_B}$ has one negative eigenvalue for $\nu>\frac{1}{3}$. The eigenvector corresponding to the negative eigenvalue is $\ket{\phi^+}=\frac{\ket{00}+\ket{11}}{\sqrt{2}}$. A linear entanglement witness can be defined to detect entangled Werner states in the following way:
	\begin{equation}
	    W_{\phi^+}=(\ket{\phi^+}\bra{\phi^+})^{T_B}. \nonumber
	\end{equation}
	By definition, it is a linear operator, but we can create a more powerful nonlinear witness operator using this,  by subtracting a nonlinear term from $W_{\phi^+}$. An example of such a nonlinear witness operator is
	\begin{equation}
	    F_{\phi^+}=\braket{W_{\phi^+}}-\frac{1}{s(X)}\left<X^{T_B}\right>\left<\left(X^{T_B}\right)^\dagger\right>. \nonumber
	\end{equation}
	Here we choose $X=\ket{\phi^+}\bra{\phi^-}$, where $\ket{\phi^-}=\frac{\ket{00}-\ket{11}}{\sqrt{2}}$. But this is an arbitrary choice. $F_{\phi^+}$ would have been a valid nonlinear witness for any other operator, $X$. It is possible to decompose $X^{T_B}$ in terms of the  hermitian 
	and anti-hermitian  
	operators,
	\begin{equation*}
	    H_{\phi^+}=\frac{1}{2}\left(\begin{matrix}
	    1&0&0&0\\
	    0&0&0&0\\
	    0&0&0&0\\
	    0&0&0&-1
	    \end{matrix}\right)\text{, and }  iA_{\phi^+}=\frac{1}{2}\left(\begin{matrix}
	    0&0&0&0\\
	    0&0&-1&0\\
	    0&1&0&0\\
	    0&0&0&0
	    \end{matrix}\right).
	\end{equation*}
	We can introduce a corresponding MDI-NEW using the hermitian matrices $W_{\phi^+}$, $H_{\phi^+}$ and $A_{\phi^+}$. The set of states
	\begin{equation}
	    \tau_0=\frac{I_2+\sigma_x}{2}\text{, }\tau_1=\frac{I_2+\sigma_y}{2}\text{, }\tau_2=\frac{I_2+\sigma_z}{2}\text{, and }\tau_3=\frac{I_2}{2},\label{eq7}
	\end{equation}
	forms a basis for
	hermitian operators on \(\mathbb{C}^2\), where $I_2$ is the identity operator on the two-dimensional Hilbert space. Thus it is possible to decompose $W_{\phi^+}$, $H_{\phi^+}$ and $A_{\phi^+}$ in terms of $\tau_s\otimes \tau_t$ 
	 as
	    \begin{eqnarray*}
		W_{\phi^+}&=&\sum_{s,t}\alpha_{st}\tau_s^T\otimes\tau_t^T,\\
		H_{\phi^+}&=&\sum_{s,t}\beta_{st}\tau_s^T\otimes\tau_t^T,\\
		\text{and }A_{\phi^+}&=&\sum_{s,t}\gamma_{st}\tau_s^T\otimes\tau_t^T,
	\end{eqnarray*} 
where the values of the real coefficients, $\alpha_{st}$, $\beta_{st}$ and $\gamma_{st}$ are expressed in the form of the following matrices
\begin{eqnarray*}
\alpha=&&\left(
\begin{matrix}
     1&0&0&-1\\
     0&1&0&-1\\
     0&0&1&-1\\
     -1&-1&-1&4
\end{matrix} \right)\text{, }\beta=\left(
\begin{matrix}
     0&0&0&0\\
     0&0&0&0\\
     0&0&0&1\\
     0&0&1&-2
\end{matrix} \right)\text{, and}\\
\gamma=&&\left(
\begin{matrix}
     0&1&0&-1\\
     -1&0&0&1\\
     0&0&0&0\\
     1&-1&0&0
\end{matrix} \right).
\end{eqnarray*}
Thus using the states given in \eqref{eq7} one can construct an MDI-NEW, that detects the same entangled states detected by \(F_{\phi^+}\).

	\section{MDI  nonlinear entanglement witnesses for PPT bound entangled states}\label{sec4}
	The states which remain positive under the action of partial transposition are called positive partial transpose or PPT states. All separable states are PPT states. Additionally, there exist some entangled states which are also PPT states. We cannot use a witness operator of the type given in Eq. \eqref{eq5} to detect PPT entangled states. In this section, we will first briefly discuss about the linear and nonlinear witness operators which can detect PPT entangled states. Finally, we will present a corresponding MDI-NEW for detection of PPT entangled states. 
	
	Let $M$ be a positive map acting on the space of operators on a Hilbert space, $\mathcal{H}_B$, and satisfying $[M(O)]^\dagger=M(O^\dagger)$, where $O$ is a positive operator. This map will also be completely positive if any extended map, $I_A\otimes M$, acting on the operators of the joint Hilbert space $\mathcal{H}_A\otimes\mathcal{H}_B$,  also preserves positivity for arbitrary dimension of subsystem $A$.
	
	Separable states on the Hilbert space $\mathcal{H}_A\otimes\mathcal{H}_B$ will remain positive under the action of $I_A\otimes M$, for any positive map $M$. Whereas, corresponding to every entangled state, there exists at least one positive map which is not completely positive and operation of which makes the entangled state go to a state with a negative eigenvalue \cite{Hor'96}. Consider an entangled state, $\rho_\xi$, on the Hilbert space $\mathcal{H}_A\otimes\mathcal{H}_B$, which acquires a negative eigenvalue under the action of the map $I_A\otimes M_\xi$. Here, $M_\xi$ is a positive map that is not completely positive. Then a corresponding (linear) witness operator can be constructed \cite{Guhne'09} as $W_\xi=(I\otimes M_\xi)^+|\xi\rangle\langle\xi|$, where $(I\otimes M_\xi)^+$ is the adjoint, of the operator $(I\otimes M_\xi)$, defined as Tr$((I\otimes M_\xi)^+ PQ)=$Tr$(P(I\otimes M_\xi)Q)$ for all $P$ and $Q$ on $\mathcal{H}_A\otimes\mathcal{H}_B$. $W_\xi$ can detect the entangled state $\rho_\xi$, which may not be an NPPT state. A corresponding NEW can be defined for detection of $\rho_\xi$ as \cite{Sen'19}
	\begin{eqnarray*}
		F_\xi(\rho)&=&\left<W_\xi \right>-\frac{1}{s(Y)g}\left< (I_A\otimes M_\xi)^+Y\right>\left<\left((I_A\otimes M_\xi)^+Y\right)^\dagger\right>\\
		&\equiv&\left<W_\xi \right>-\frac{1}{s(Y)g}\langle \widetilde{Y}\rangle\langle\widetilde{Y}^\dagger\rangle.
	\end{eqnarray*}
	Here $Y=|\xi\rangle\langle\zeta|$, $|\zeta\rangle$ is any arbitrary state, and $s(Y)$ is the square of the largest Schmidt coefficient of the state $|\zeta\rangle$. The constant $g$ is given by $g=\max_\sigma$Tr$[M(\sigma)]$, where the maximization is taken over the entire state space. The $\widetilde{Y}$ in the above equation can be written as a sum of  hermitian and antihermitian operators, say as $\widetilde{Y}=H_\xi+iA_\xi$. We can decompose the hermitian operators $W_\xi$, $H_\xi$, and $A_\xi$ in the following way:
	\begin{eqnarray*}
	    	W_\xi&=&\sum_{s,t}\lambda_{st}\pi_s^T\otimes\chi_t^T, \quad
		H_\xi=\sum_{s,t}\mu_{st}\pi_s^T\otimes\chi_t^T, \\
		A_\xi&=&\sum_{s,t}\nu_{st}\pi_s^T\otimes\chi_t^T,	
	\end{eqnarray*}
	where $\pi_s$ and $\chi_t$ are density operators acting on the Hilbert spaces $\mathcal{H}_A$ and $\mathcal{H}_B$ respectively, and $\lambda_{st}$, $\mu_{st}$, and $\nu_{st}$ are real numbers.
	Using similar algebra and strategies as in the preceding section,
	it is possible to utilize the 
	function,
	\begin{widetext}
	\begin{equation}
	N_\xi(P)=\sum_{s,t}\lambda_{s,t}P(1,1|\pi_s,\chi_t)-\frac{1}{s(Y)g d_A d_B P(1,1|m_A,m_B)}
	\left[\left(\sum_{s,t}\mu_{s,t}P(1,1|\pi_s,\chi_t)\right)^2+\left(\sum_{s,t}\nu_{s,t}P(1,1|\pi_s,\chi_t)\right)^2\right], 
	\end{equation}
	\end{widetext}
	as an MDI-NEW. 
	However, unlike the MDI-NEW in the preceding section, 
$N_\xi(P_\rho)$ can be used to  detect the state $\rho_\xi$ in a measurement-device-independent way, when the 
measurement operator corresponding to outcome 1 at Alice (Bob) is $\ket{\Phi^+_{A(B)}}\bra {\Phi^+_{A(B)}}$.

\vspace{2 mm}

\noindent \textbf{Example:} The set of states
\begin{equation*}
    \rho_B=\frac{2}{7}\ket{\tilde{\psi}}\bra{\tilde{\psi}}+\frac{a}{7}\sigma_++\frac{5-a}{7}\sigma_-,
\end{equation*}
where
\begin{eqnarray*}
    \ket{\tilde{\psi}}&=&\frac{1}{\sqrt{3}}(\ket{00}+\ket{11}+\ket{22}),\\
    \sigma_+&=&\frac{1}{3}(\ket{01}\bra{01}+\ket{12}\bra{12}+\ket{20}\bra{20}),\\
    \sigma_-&=&\frac{1}{3}(\ket{10}\bra{10}+\ket{21}\bra{21}+\ket{02}\bra{02}),
\end{eqnarray*}
are PPT for $a\leq 4$ \cite{PRL} But if one partially applies the following map \cite{choi}
\begin{eqnarray*}
    M_1\left(\left[\begin{matrix}
         a_{11}&a_{12}&a_{13}\\
         a_{21}&a_{22}&a_{23}\\
         a_{31}&a_{32}&a_{33}
    \end{matrix}\right]\right)=\left[\begin{matrix}
         a_{11}+a_{33}&-a_{12}&-a_{13}\\
         -a_{21}&a_{22}+a_{11}&-a_{23}\\
         -a_{31}&-a_{32}&a_{33}+a_{22}
    \end{matrix}\right]
\end{eqnarray*}
on the state $\rho_B$, the resulting operator, i.e., $I_3\otimes M_1(\rho_B)$, becomes negative for $3<a\leq4$, where $I_3$ is the identity operator on the three-dimensional Hilbert space. Thus, the state $\rho_B$ is at the same time PPT and entangled, and so bound entangled in the range $3<a\leq4$. The eigen-vector of $I_3\otimes M_1(\rho_B)$ giving rise to a negative eigenvalue is $\ket{\tilde{{\psi}}}$. Thus the witness operator, $W_{\tilde{\psi}}=(I_3\otimes M_1)^+(\ket{\tilde{{\psi}}}\bra{\tilde{{\psi}}})$, can detect the bound entangled states $\rho_B$, in the range $3<a\leq4$. A corresponding nonlinear witness operator can be defined as
\begin{equation}
    F_{\tilde{\psi}}=\braket{W_{\tilde{\psi}}}-\frac{1}{s(Y)g}\braket{(I_3\otimes M_1)^+(Y)}\braket{(I_3\otimes M_1)^+ (Y))^\dagger}.
\end{equation}
In this case, $g=2$. We take $Y=\frac{1}{2\sqrt{3}}(\ket{00}+\ket{11}+\ket{22})(\bra{01}+\bra{10}+\bra{12}+\bra{21})$. We can decompose $(I_3\otimes M_1)^+Y$ in terms of the hermitian matrix, $H_{\tilde{\psi}}$, and the anti-hermitian matrix $iA_{\tilde{\psi}}$. Now to construct the MDI-NEW, we have to define a set of basis elements for $3\times 3 $ hermitian operators. We consider the basis with the elements,
\begin{equation}
    \pi_0=\frac{I_3}{3}\text{, }\pi_s=\frac{I_3+\Lambda_s}{3},
    \text{ and } \pi_8=\frac{I_3+\frac{\sqrt{3}}{2}\Lambda_8}{3},\label{eqe}
\end{equation}
where $s$ runs from 1 to 7. The $\Lambda_s$'s and $\Lambda_8$ are the Gell-Mann matrices. The hermitian matrices, $W_{\tilde{\psi}}$, $H_{\tilde{\psi}}$, and $A_{\tilde{\psi}}$ can be written in terms of the basis elements, $\pi_s\otimes\pi_t$, for \(s,t \in \{0,1,2,\ldots,8\}\), as
	\begin{eqnarray*}
		W_{\tilde{\psi}}=\sum_{s,t}\lambda_{st}\pi_s^T\otimes\pi_t^T,\\
		H_{\tilde{\psi}}=\sum_{s,t}\mu_{st}\pi_s^T\otimes\pi_t^T,\\
		A_{\tilde{\psi}}=\sum_{s,t}\nu_{st}\pi_s^T\otimes\pi_t^T.	
	\end{eqnarray*}
The coefficients $\lambda_{st}$, $\mu_{st}$, and $\nu_{st}$ are found to be given by
\begin{widetext}
\begin{eqnarray*}
    \lambda&=&\left(\begin{matrix}
         \frac{15}{4}&\frac{3}{2}&-\frac{3}{2}&\frac{-9}{4}&\frac{3}{2}&-\frac{3}{2}&\frac{3}{2}&-\frac{3}{2}&\frac{1}{2}\\
         \frac{3}{2}&-\frac{3}{2}&0&0&0&0&0&0&0\\
         -\frac{3}{2}&0&\frac{3}{2}&0&0&0&0&0&0\\
         \frac{3}{4}&0&0&\frac{3}{4}&0&0&0&0&-\frac{3}{2}\\
         \frac{3}{2}&0&0&0&-\frac{3}{2}&0&0&0&0\\
         -\frac{3}{2}&0&0&0&0&\frac{3}{2}&0&0&0\\
         \frac{3}{2}&0&0&0&0&0&-\frac{3}{2}&0&0\\
         -\frac{3}{2}&0&0&0&0&0&0&\frac{3}{2}&0\\
         -\frac{5}{2}&0&0&\frac{3}{2}&0&0&0&0&1
    \end{matrix}\right), \quad
    \mu=\left(\begin{matrix}
        -\frac{3\sqrt{3}}{8}&\frac{3\sqrt{3}}{8}&-\frac{3\sqrt{3}}{8}&0&\frac{3\sqrt{3}}{4}&-\frac{3\sqrt{3}}{4}&-\frac{3\sqrt{3}}{4}&-\frac{3\sqrt{3}}{8}&
        -\frac{\sqrt{3}}{2}\\
        \frac{3\sqrt{3}}{4}&0&0&\frac{3\sqrt{3}}{8}&-\frac{3\sqrt{3}}{8}&0&0&0&\frac{\sqrt{3}}{4}\\
        -\frac{3\sqrt{3}}{8}&0&0&0&0&\frac{3\sqrt{3}}{8}&0&0&0\\
        -\frac{3\sqrt{3}}{8}&0&0&0&0&0&\frac{3\sqrt{3}}{8}&0&0\\
        \frac{3\sqrt{3}}{4}&-\frac{3\sqrt{3}}{8}&0&0&0&0&-\frac{3\sqrt{3}}{8}&0&0\\
        -\frac{3\sqrt{3}}{4}&0&\frac{3\sqrt{3}}{8}&0&0&0&0&\frac{3\sqrt{3}}{8}&0\\
        \frac{3\sqrt{3}}{2}&0&0&-\frac{3\sqrt{3}}{8}&-\frac{3\sqrt{3}}{8}&0&0&0&\frac{\sqrt{3}}{4}\\
        -\frac{3\sqrt{3}}{8}&0&0&0&0&\frac{3\sqrt{3}}{8}&0&0&0\\
        \frac{\sqrt{3}}{4}&-\frac{\sqrt{3}}{2}&0&0&0&0&\frac{\sqrt{3}}{4}&0&0
    \end{matrix}\right),\\ \\ \\
    \nu&=&\left(\begin{matrix}
        -\frac{3\sqrt{3}}{8}&\frac{3\sqrt{3}}{8}&\frac{9\sqrt{3}}{8}&-\frac{3\sqrt{3}}{4}&0&0&-\frac{3\sqrt{3}}{8}&0&0\\
        \frac{3\sqrt{3}}{8}&0&0&0&0&-\frac{3\sqrt{3}}{8}&0&0&0\\
        \frac{3\sqrt{3}}{4}&0&0&\frac{3\sqrt{3}}{8}&-\frac{3\sqrt{3}}{8}&0&0&0&-\frac{3\sqrt{3}}{4}\\
        \frac{3\sqrt{3}}{8}&0&-\frac{3\sqrt{3}}{4}&0&0&0&0&\frac{3\sqrt{3}}{8}&0\\
        0&0&-\frac{3\sqrt{3}}{8}&0&0&0&0&\frac{3\sqrt{3}}{8}&0\\
        0&-\frac{3\sqrt{3}}{8}&0&0&0&0&\frac{3\sqrt{3}}{8}&0&0\\
        -\frac{3\sqrt{3}}{8}&0&0&0&0&\frac{3\sqrt{3}}{8}&0&0&0\\
        -\frac{3\sqrt{3}}{2}&0&0&\frac{3\sqrt{3}}{8}&\frac{3\sqrt{3}}{8}&0&0&0&\frac{3\sqrt{3}}{4}\\
        \frac{3\sqrt{3}}{4}&0&0&0&0&0&0&-\frac{3\sqrt{3}}{4}&0
    \end{matrix}\right).
    \end{eqnarray*}
\end{widetext}
Hence, using the basis elements given in Eqs.~\eqref{eqe}, one can construct an MDI-NEW that detects the same states as those detected by \(F_{\tilde{\psi}}\). In particular, it can detect the bound entangled states in the class, \(\rho_B\).

	\section{Conclusion}
	Using EWs, one can efficiently detect entangled states by measuring the expectation value of the witness operator. But in experiments, measurements cannot be done with a cent percent 
	accuracy. To avoid the misidentification of a separable state as an entangled one, 
	that can arise from the errors present in the implementation of the measurement operators, MDI-EWs were introduced based on linear EWs. 
	On the other hand, NEWs have been introduced, that can detect a strictly larger set of entangled states than the corresponding EWs. 
	Here we have presented an entanglement witness with both the properties, viz. measuremnt-device independence and nonlinearity, and named them as MDI-NEWs. We have shown that
	independent of the measurements being performed, MDI-NEWs will never wrongly certify a separable state as entangled. Furthermore, 
	we have shown that the MDI-NEWs can detect all the entangled states that can be detected using corresponding NEWs, therefore performing better than the (linear) EWs and MDI-EWs. We have first derived MDI-NEWs for detection of NPPT states. Subsequently, we have  also provided MDI-NEWs for detection of PPT entangled states. Though we have considered two particular forms of NEWs to derive the corresponding MDI-NEWs, the method presented can be utilized for creating a MDI-NEW corresponding to any other NEW.  
	\label{sec5}

\begin{acknowledgments}
The research of CS was supported partly by the INFOSYS scholarship.
We acknowledge support from the Department of Science and Technology, Government of India through the QuEST grant (grant number DST/ICPS/QUST/Theme-3/2019/120). 
\end{acknowledgments}

\end{document}